\documentclass[floats,floatfix,twocolumn,superscriptaddress,nofootinbib,showpacs]{revtex4-1}

\pdfoutput=1
\usepackage{amsfonts,amssymb,amsmath}
\usepackage[usenames,dvipsnames]{xcolor}
\usepackage[unicode,colorlinks=true,linkcolor=blue,urlcolor=blue,citecolor=gray]{hyperref}
\usepackage[all]{hypcap}
\usepackage{graphicx}
\usepackage{xspace}
\usepackage{tikz}
\usetikzlibrary{shapes,arrows}
\usetikzlibrary{matrix}
\usetikzlibrary{shapes.multipart}
\usetikzlibrary{er,positioning}
\usepackage[normalem]{ulem} %for \sout
\usepackage{url}
\usepackage{color,soul}
\usepackage[caption=false]{subfig}
\usepackage[T1]{fontenc}
\usepackage[utf8]{inputenc}
\usepackage{microtype}
\newcommand{\Flatiron}{\affiliation{Center for Computational Astrophysics, Flatiron Institute, 162 5th Ave, New York, NY 10010}}

\newcommand{\eps}{\ensuremath{\varepsilon}}

\newcommand{\pd}{\partial}

\newcommand{\nn}{\nonumber}

\newcommand{\txt}[1]{{\textrm{\tiny{#1}}}}
\newcommand{\mpl}{\ensuremath{m_\txt{pl}}}

\newcommand{\dual}{\,{}^*\!}  
\newcommand{\agb}{\alpha_\mathrm{GB}}
\newcommand{\sagb}{\sqrt{\alpha_\mathrm{GB}}}
\newcommand{\sgmagb}{\sqrt{\alpha_\mathrm{GB}}/GM}

\begin{document}

\title{Numerical relativity simulation of GW150914 in Einstein dilaton Gauss-Bonnet gravity}

\author{Maria Okounkova}
\Flatiron
\email{mokounkova@flatironinstitute.org}

\date{\today}

\begin{abstract}
A present challenge in testing general relativity (GR) with binary black hole gravitational wave detections is the inability to perform model-dependent tests due to the lack of merger waveforms in beyond-GR theories. In this study, we produce the first numerical relativity binary black hole gravitational waveform in Einstein dilaton Gauss-Bonnet (EDGB) gravity, a higher-curvature theory of gravity with motivations in string theory. We evolve a binary black hole system in order-reduced EDGB gravity, with parameters consistent with GW150914. We focus on the merger portion of the waveform, due to the presence of secular growth in the inspiral phase. We compute mismatches with the corresponding general relativity merger waveform, finding that from a post-inspiral-only analysis, we can constrain the EDGB lengthscale to be $\sagb \lesssim 11$ km. 
\end{abstract}

\maketitle

\section{Introduction}

Though Einstein's theory of general relativity (GR) has passed all precision tests to date, at some lengthscale, it must break down and be reconciled with quantum mechanics in a beyond-GR theory of gravity. Binary black hole (BBH) mergers probe the strong-field, non-linear regime of gravity, and thus gravitational wave signals from these systems could contain signatures of a beyond-GR theory. While LIGO presently performs \textit{model-independent} and \textit{parametrized} tests of general relativity~\cite{TheLIGOScientific:2016src, LIGOScientific:2019fpa}, one important additional avenue of looking for deviations from general relativity is to perform \textit{model-dependent} tests. Such model-dependent tests require access to numerical waveforms in beyond-GR theories of gravity through merger, the lack of which is currently a severe limitation on constraining beyond-GR physics~\cite{Yunes:2016jcc}. 

We produce the first numerical relativity gravitational waveforms in Einstein dilaton Gauss-Bonnet (EDGB) gravity, an effective field theory that modifies the Einstein-Hilbert action of GR through the inclusion of a scalar field coupled to terms quadratic in curvature. These terms are meant to encompass underlying quantum gravity effects, in particular motivated by string theory~\cite{Kanti:1995vq, 1987NuPhB.291...41G, Moura:2006pz, Berti:2015itd}, and the coupling to the scalar field is governed by an EDGB lengthscale parameter $\sagb$. The well-posedness of the initial value problem in full EDGB gravity is unknown~\cite{Papallo:2017qvl, Papallo:2017ddx, Ripley:2019hxt, Ripley:2019irj}. We thus work in an \textit{order-reduction scheme}, in which we perturb the EDGB scalar field and spacetime metric about a GR background. 

Previously, Witek et al.~\cite{Witek:2018dmd} evolved the leading-order EDGB scalar field on a BBH background, predicting a bound of $\sagb \lesssim 2.7$ km on the EDGB lengthscale, a constraint seven orders of magnitude tighter than observational results from solar-system tests. In this study, we evolve the leading-order EDGB correction to the spacetime metric on a BBH background, thus obtaining the \textit{leading-order EDGB modification to the merger gravitational waveform}. We compute mismatches between the GR and EDGB-corrected waveforms, aiming to similarly bound the EDGB lengthscale. 

We focus on an astrophysically-relevant BBH system with spin and mass ratio consistent with GW150914, the loudest LIGO detection to date~\cite{Abbott:2016blz, TheLIGOScientific:2016wfe, LIGOScientific:2018mvr}, for which significant model-independent and parametrized tests of GR have been performed~\cite{TheLIGOScientific:2016src, Yunes:2016jcc, LIGOScientific:2019fpa, Isi:2019aib, Nair:2019iur}. This extends our results in~\cite{MashadCSGWPaper}, where we simulated the same system in dynamical Chern-Simons gravity (dCS), another quadratic beyond-GR theory with motivations in string theory and loop quantum gravity~\cite{Alexander:2009tp, Green:1984sg, Taveras:2008yf, Mercuri:2009zt}.

\section{Setup}
\label{sec:setup}

We set $G = c = 1$ throughout.  Quantities are given in terms of units of $M$, the sum of the Christodolou masses of the background black holes at a given reference time~\cite{Boyle:2009vi}. Latin letters in the beginning of the alphabet $\{a, b,  c, d \ldots \}$ denote 4-dimensional spacetime indices, and $g_{ab}$ refers to the spacetime metric with covariant derivative $\nabla_c$.

\subsection{Equations of motion}
\label{sec:equations}

The overall form of the EDGB action that we will use in this paper, chosen to be consistent with Witek et al.~\cite{Witek:2018dmd}, is

\begin{align}
\label{eq:action}
    S \equiv \int \frac{\mpl^2}{2} d^4 x \sqrt{-g} \left[R - \frac{1}{2} (\pd \vartheta)^2 + 2 \alpha_\mathrm{GB} f(\vartheta) \mathcal{R}_\mathrm{GB} \right]\,,
\end{align}
where the first term is the  Einstein-Hilbert action of GR (where $R$ is the 4-dimensional Ricci scalar), $\vartheta$ is the EDGB scalar field, and $\alpha_\mathrm{GB}$ is the EDGB coupling parameter with dimensions of length squared. We will work with $\sagb$, which has dimensions of length, throughout this paper. Finally, $\mathcal{R}_\mathrm{GB}$ is the EDGB scalar, of the form 
\begin{align}
\label{eq:RGBDefinition}
    \mathcal{R}_\mathrm{GB} = R^{abcd}R_{abcd} - 4 R^{ab} R_{ab} + R^2\,.
\end{align}

It is unknown whether EDGB has a well-posed initial value problem~\cite{Papallo:2017qvl, Papallo:2017ddx, Ripley:2019hxt, Ripley:2019irj}. However, as we have done in~\cite{Okounkova:2017yby, MashaEvPaper, MashaHeadOn, MashadCSGWPaper}, we perturb the spacetime metric and EDGB scalar field about an arbitrary GR background as
\begin{align}
    g_{ab} &= g_{ab}^{(0)} + \sum_{n = 1}^{\infty} \eps^n g_{ab}^{(n)}\,, \\
    \vartheta &= \sum_{n = 0}^\infty \eps^n \vartheta^{(n)}\,,
\end{align}
where $\eps$ is an order-counting parameter that counts powers of $\agb$, and superscript ${}^{(0)}$ corresponds to the GR solution, which we refer to as \textit{the background}. 

At each order, the equations of motion are well-posed. Moreover, the EDGB coupling parameter $\agb$ scales out at each order, and thus we only need to perform \textit{one} BBH simulation for each set of GR background parameters. 

Zeroth order corresponds to pure general relativity. The equation of motion for the zeroth order scalar field, $\vartheta^{(0)}$, corresponds to a scalar field minimally coupled to vacuum GR, and thus $\vartheta^{(0)}$ should decay to zero in BH spacetimes by the no-hair theorem. 

The leading-order EDGB scalar field appears at first-order as $\vartheta^{(1)}$, sourced by the curvature of the GR background,  with equation of motion (cf.~\cite{Witek:2018dmd} for a full derivation),
\begin{align}
    \square^{(0)} \vartheta^{(1)} &= - M^2 \mathcal{R}^{(0)}_\mathrm{GB}\,, \\
    \label{eq:RGB0Definition}
     \mathcal{R}^{(0)}_\mathrm{GB} &\equiv R^{(0)} {}^{abcd}R^{(0)}_{abcd} - 4 R^{(0)} {}^{ab} R^{(0)}_{ab} + R^{(0)} {}^2\,,
\end{align}
where the superscript ${}^{(0)}$ refers to quantities computed from the GR background. Here, the leading-order correction to $f(\vartheta)$ has been set to $\frac{1}{8}$, in accordance with~\cite{Witek:2018dmd}. 

Meanwhile, the leading EDGB deformation to the spacetime metric comes in at \textit{second} order, with the equation of motion (cf.~\cite{Witek:2018dmd}),
\begin{align}
\label{eq:DeltapsiEOM}
    G_{ab}^{(0)} [g_{ab}^{(2)}] = - 8 M^2 \mathcal{G}_{ab}^{(0)}[\vartheta^{(1)}] + T_{ab}[\vartheta^{(1)}]\,.
\end{align}

In the above equations, $T_{ab}[\vartheta^{(1)}]$ is the standard Klein-Gordon stress energy tensor associated with $\vartheta^{(1)}$, of the form\footnote{Note that our definition of $T_{ab}[\vartheta^{(1)}]$ in Eq.~\eqref{eq:Tab1} differs from Eq. 15 in~\cite{Witek:2018dmd} by a factor of $2$, and hence the $T_{ab}[\vartheta^{(1)}]$ term in Eq.~\eqref{eq:DeltapsiEOM} differs by a factor of $2$. We have chosen this convention to be in line with the canonical form of the Klein-Gordon stress-energy tensor.}
\begin{align}
\label{eq:Tab1}
    T_{ab}[\vartheta^{(1)}] = \nabla^{(0)}_a \vartheta^{(1)} \nabla^{(0)}_b \vartheta^{(1)} - \frac{1}{2} g_{ab}^{(0)} \nabla^{(0)}_c \vartheta^{(1)} \nabla^{(0)}{}^c \vartheta^{(1)}\,,
\end{align}
and 
\begin{align}
\label{eq:G0Definition}
    \mathcal{G}_{ab}^{(0)}[\vartheta^{(1)}] = 2 \epsilon^{edfg} g_{c(a}^{(0)} g_{b)d}^{(0)} \nabla_h^{(0)} \left[\frac{1}{8} \dual R^{(0)}{}^{ch} {}_{fg} \nabla^{(0)}_e \vartheta^{(1)} \right]\,,
\end{align}
where $\dual R^{ab} {}_{cd} = \epsilon^{abef} R_{efcd}^{(0)}$ and $\epsilon^{abcd}$ is the Levi-Citiva pseudo-tensor, with $\epsilon^{abcd} = -[abcd]/\sqrt{-g^{(0)}}$, where $[abcd]$ is the alternating symbol.  

Note that we work on a vacuum GR background, and thus terms vanish to give the simplified equations of motion 
\begin{align}
\label{eq:ThetaEvolution}
    \square^{(0)} \vartheta^{(1)} &= - M^2 R^{(0)} {}^{abcd}R^{(0)}_{abcd} \\
    \label{eq:MetricEvolution}
    \nn G_{ab}^{(0)} [g_{ab}^{(2)}] &= - 
    2 M^2 \epsilon^{edfg} g_{c(a}^{(0)} g_{b)d}^{(0)} \dual R^{(0)}{}^{ch} {}_{fg}\nabla_h^{(0)}  \nabla^{(0)}_e \vartheta^{(1)} \\
    & \quad \quad  \quad  \quad + T_{ab}[\vartheta^{(1)}]\,.
\end{align}

\textbf{To summarize:} the order-reduction procedure is illustrated in Fig. 1 of~\cite{MashadCSGWPaper}. We will have a GR binary black hole background. The curvature of this background will then source the leading-order EDGB scalar field (Eq.~\eqref{eq:ThetaEvolution}). This leading-order scalar field and the GR background will then source the leading-order EDGB correction to the spacetime (Eq.~\eqref{eq:MetricEvolution}), which in turn will give us the leading-order EDGB correction to the gravitational waveform.

\subsection{Secular growth during inspiral}
\label{sec:secular_growth}

As we initially noted in~\cite{MashadCSGWPaper}, the perturbative order-reduction scheme outlined in Sec.~\ref{sec:equations} gives rise to secular growth during the inspiral. In the order-reduction scheme, the rate of inspiral is governed by the GR background. However, in the full, non-linear EDGB theory, we expect the black holes to have a faster rate of inspiral due to energy loss to the scalar field~\cite{Yagi:2011xp}. Since we do not backreact on the GR background in the order-reduction scheme, we do not capture this correction to the rate of inspiral, and hence our solution contains secular growth. This is a feature generically found in perturbative treatments~\cite{MR538168}, including in extreme mass-ratio inspirals~\cite{Hinderer:2008dm}.

When we simulated an inspiraling binary black hole system in order-reduced dCS~\cite{MashadCSGWPaper}, we indeed observed secular growth during the inspiral. We performed a set of simulations where we \textit{ramped on} the dCS source terms at various start times during the inspiral, for the same set of background parameters. We found secular growth in the amplitude of the resulting dCS correction to the waveform, with simulations with earlier start times having larger amplitudes. \textit{However}, this secular growth settled to a quadratic minimum for a start time before the portion of the inspiral-merger present in the LIGO band for a GW150914-like system. Thus, we were able to focus on this portion of the waveform in~\cite{MashadCSGWPaper} without having contamination from secular effects. 

In this study, we apply the same procedure, where we ramp on the EDGB source terms at a variety of start times for the same (long) GR binary black hole background simulation. We search for the start time at which the waveform is no longer contaminated by secular effects, and present the resulting merger waveform. 

The inspiral in EDGB is more strongly modified from GR, with the modifications to the inspiral occurring at -1 PN order relative to GR due to the presence of dipolar radiation in the scalar field~\cite{Yagi:2011xp}. This is 3 PN orders higher than the leading modification in the dCS case, where dipolar radiation is absent during inspiral. Thus we expect the minimum of the secular growth to occur later in the inspiral in EDGB than in dCS for the same physical system.

\subsection{Computational details} 

Eqs.~\eqref{eq:ThetaEvolution} and~\eqref{eq:MetricEvolution} are precisely the equations that we co-evolve with the GR background. We use the Spectral Einstein Code~\cite{SpECwebsite}, which uses pseudo-spectral methods and thus guarantees exponential convergence in the fields. All of the technical details are given in~\cite{MashadCSGWPaper, MashaHeadOn, MashaEvPaper, Okounkova:2017yby}. The domain decomposition is precisely that of the analogous dCS study~\cite{MashadCSGWPaper}.

\section{EDGB merger waveforms}
\label{sec:Results}

\subsection{Simulation parameters}
\label{sec:simulation}

While there is a distribution of mass and spin parameters consistent
with GW150914~\cite{TheLIGOScientific:2016wfe, Kumar:2018hml}, we
choose to use the parameters of SXS:BBH:0305, as given in the
Simulating eXtreme Spacetimes (SXS) catalog~\cite{SXSCatalog}. This
simulation was used in Fig. 1 of the GW150914 detection
paper~\cite{Abbott:2016blz}, as well a host of follow-up
studies~\cite{Lovelace:2016uwp, Bhagwat:2017tkm, Giesler:2019uxc}. We additionally used precisely these parameters for our dCS BBH simulation~\cite{MashadCSGWPaper}. The
configuration has initial dimensionless spins $\chi_A = 0.330 \hat{z}$
and $\chi_B = -0.440 \hat{z}$, aligned and anti-aligned with the
orbital angular momentum. The dominant GR spherical harmonic modes of
the gravitational radiation for this system are $(l,m)=(2,\pm2)$. The
system has initial masses of $0.5497 \,M$ and $0.4502 \,M$, leading to
a mass ratio of $1.221$. The initial eccentricity is
$\sim 8 \times 10^{-4}$. The remnant has final
Christodolou mass $0.9525 \,M$ and dimensionless spin $0.692$ purely
in the $\hat{z}$ direction. The GR background simulation completes 23 orbits before merger.

\subsection{Regime of validity}
\label{sec:validity}

The results that we present for the leading-order EDGB scalar and gravitational waveforms have the EDGB coupling parameter $\agb$ scaled out. For the perturbative order reduction scheme to be valid, we require that $g_{ab}^{(2)} \lesssim C g_{ab}^{(0)}$, for some constant $C < 1$. This in turn becomes a constraint on $\agb$, of the form (cf.~\cite{Okounkova:2017yby} for an analogous derivation)
\begin{align}
\label{eq:Validity}
    \frac{\sqrt{\agb}}{GM} \lesssim \left(C \frac{\|g_{ab}^{(0)}\|}{\|g_{ab}^{(2)}\|} \right)^{1/4}\,.
\end{align}
We choose $C = 0.1$, and evaluate Eq.~\eqref{eq:Validity} on each slice of the numerical relativity simulation. We find the the strongest constraint on the allowed value of $\sgmagb$ comes at merger, when the spacetime is most highly perturbed, with a value of $\sagb/GM \sim 0.17$ for the simulation presented in this paper.

\subsection{EDGB scalar field waveforms}
\label{sec:scalar_waveforms}
\begin{figure}
  \includegraphics[width=\columnwidth]{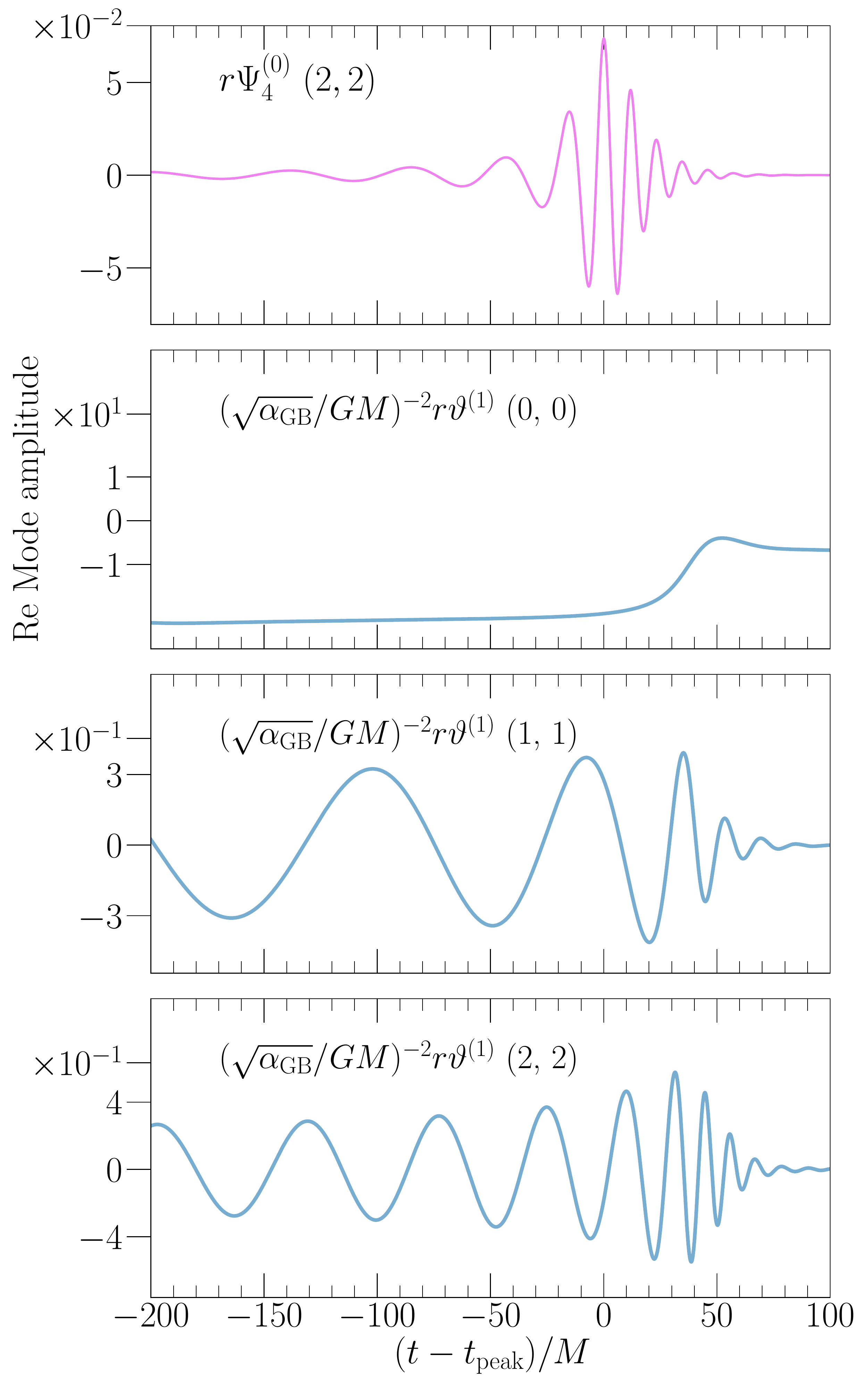}
  \caption{Dominant modes of the leading-order EDGB scalar field $\vartheta^{(1)}$, decomposed into spherical harmonics $(l, m)$, as a function of time relative to the peak time of the GR gravitational waveform. The top panel corresponds to the dominant $(2,2)$ mode of the GR gravitational radiation for comparison. The bottom three panels correspond to the dominant modes of $\vartheta^{(1)}$, which are $(l,m=l)$. We see the presence of $l = 1$ dipolar radiation during the inspiral. While the $l = 0$ monopole is non-radiative during the inspiral, we see a burst of monopolar radiation at merger. Compare with Fig. 4 of~\cite{Witek:2018dmd} and the dCS case in Fig. 1 of~\cite{Okounkova:2017yby}. Note that the $\vartheta^{(1)}$ waveforms have the EDGB coupling $\sgmagb$ scaled out, and thus an appropriate value (cf. Sec.~\ref{sec:validity}) of this coupling parameter must be re-introduced for the results to be physically meaningful. 
  }
  \label{fig:ScalarFields}
\end{figure}

In Fig.~\ref{fig:ScalarFields}, we show the results for the leading-order EDGB scalar field, $\vartheta^{(1)}$. We decompose the scalar field into spherical harmonics, and find that the dominant modes are $(l, m = l)$, in accordance with~\cite{Witek:2018dmd, Yagi:2011xp}. We see the presence of $l = 1$ dipolar radiation in the field during inspiral, in accordance with~\cite{Witek:2018dmd, Yagi:2011xp}. We see that the monopolar $l = 0$ mode is non-radiative during the inspiral, but that there is a burst of monopolar radiation at merger. This is in agreement with~\cite{Witek:2018dmd}, and moreover is similar to the results in dCS~\cite{Okounkova:2017yby}, where we found that the leading non-radiative mode (the dipole in the dCS case) exhibits a burst of radiation at merger. 

\subsection{EDGB gravitational waveforms}
\label{sec:merger_waveform}

As explained in Sec.~\ref{sec:secular_growth}, because of secular growth during the inspiral, we focus on simulations with EDGB effects ramped on close to merger, in order to mitigate the amount of secular growth from the inspiral (we give more details in Sec.~\ref{sec:observed_secular}). We thus present these merger waveforms in this section. 

From the leading-order EDGB metric deformation $ g_{ab}^{(2)}$, we can compute $\Psi_4^{(2)}$, the leading-order modification to the gravitational waveform, given by the Newman-Penrose scalar $\Psi_4$. Note that $g_{ab}^{(2)}$ and hence $\Psi_4^{(2)}$ from the simulation are independent of the EDGB coupling parameter. In order to produce a full, second-order-accurate EDGB gravitational waveform, we must add $\Psi_4^{(2)}$ to the background GR waveform $\Psi_4^{(0)}$ as
\begin{align}
\label{eq:total_psi4}
    \Psi_4 = \Psi_4^{(0)} + (\sgmagb)^4  \Psi_4^{(2)} + \mathcal{O}((\sgmagb)^6)\,,
\end{align}
for a given choice for the EDGB coupling parameter $\sgmagb$. We require that $\sgmagb$ lies within the regime of validity for the perturbative scheme as given in   Sec.~\ref{sec:validity}

\begin{figure}
  \includegraphics[width=\columnwidth]{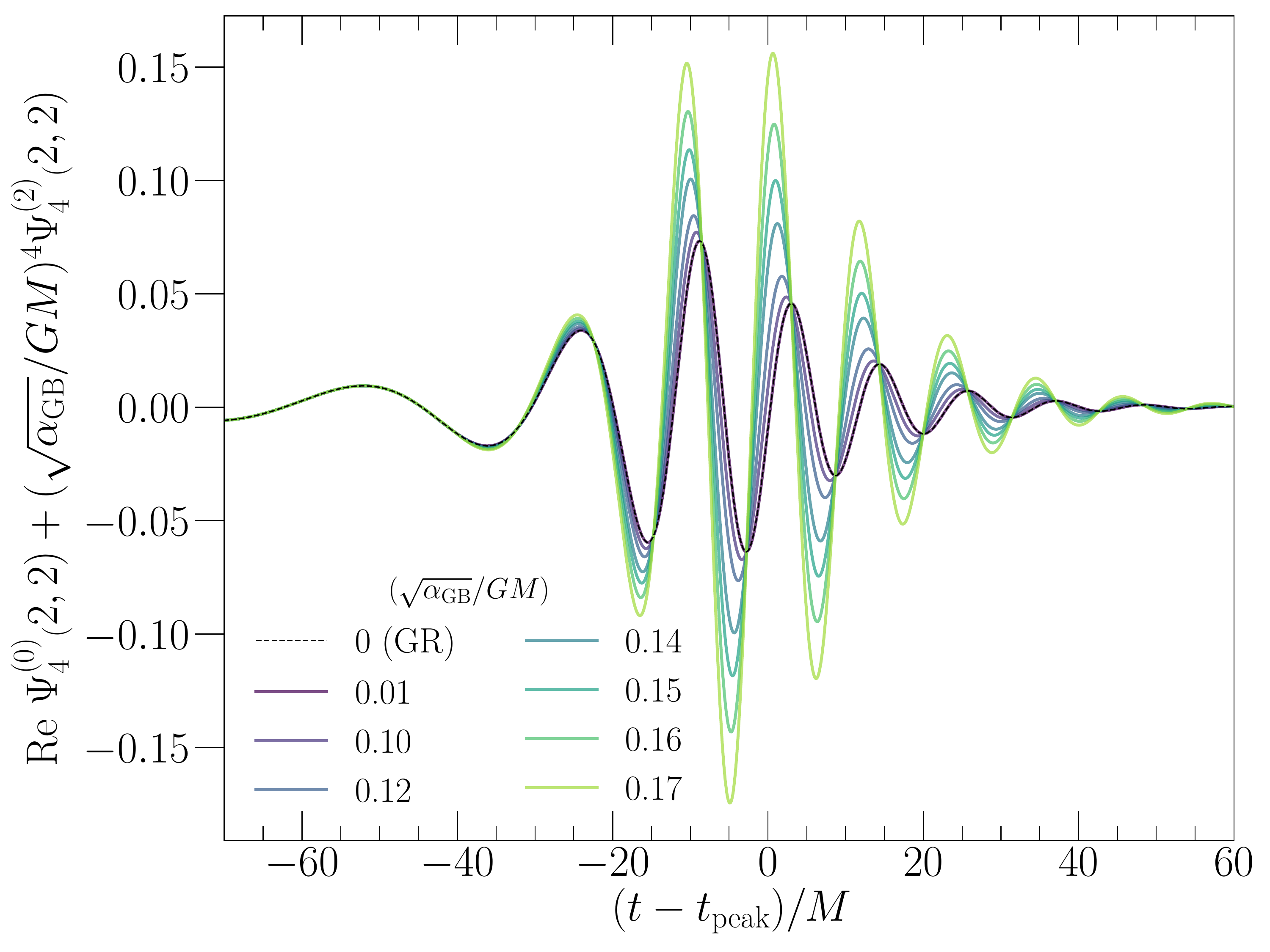}
  \caption{EDGB-corrected merger gravitational waveforms, as computed from Eq.~\eqref{eq:total_psi4}, for a variety of values of the EDGB coupling parameter $\sgmagb$. The dashed black line, with $\sgmagb = 0$, corresponds to the GR waveform. The value $\sgmagb = 0.17$ corresponds to the maximal allowed value in order for the perturbative scheme to be valid (cf. Sec.~\ref{sec:validity}). We see that the EDGB-corrected waveform has both an amplitude and phase shift relative to GR. 
  }
  \label{fig:Together}
\end{figure}

In Fig.~\ref{fig:Together}, we show this total waveform for a variety of values of $\sgmagb$. We see that the EDGB-corrected waveform has an amplitude shift relative to GR, as well as a phase shift, consistent with the notion that EDGB should have a faster inspiral due to energy loss to the scalar field~\cite{Yagi:2011xp}. 

\subsection{Secular growth}
\label{sec:observed_secular}

\begin{figure}
  \includegraphics[width=\columnwidth]{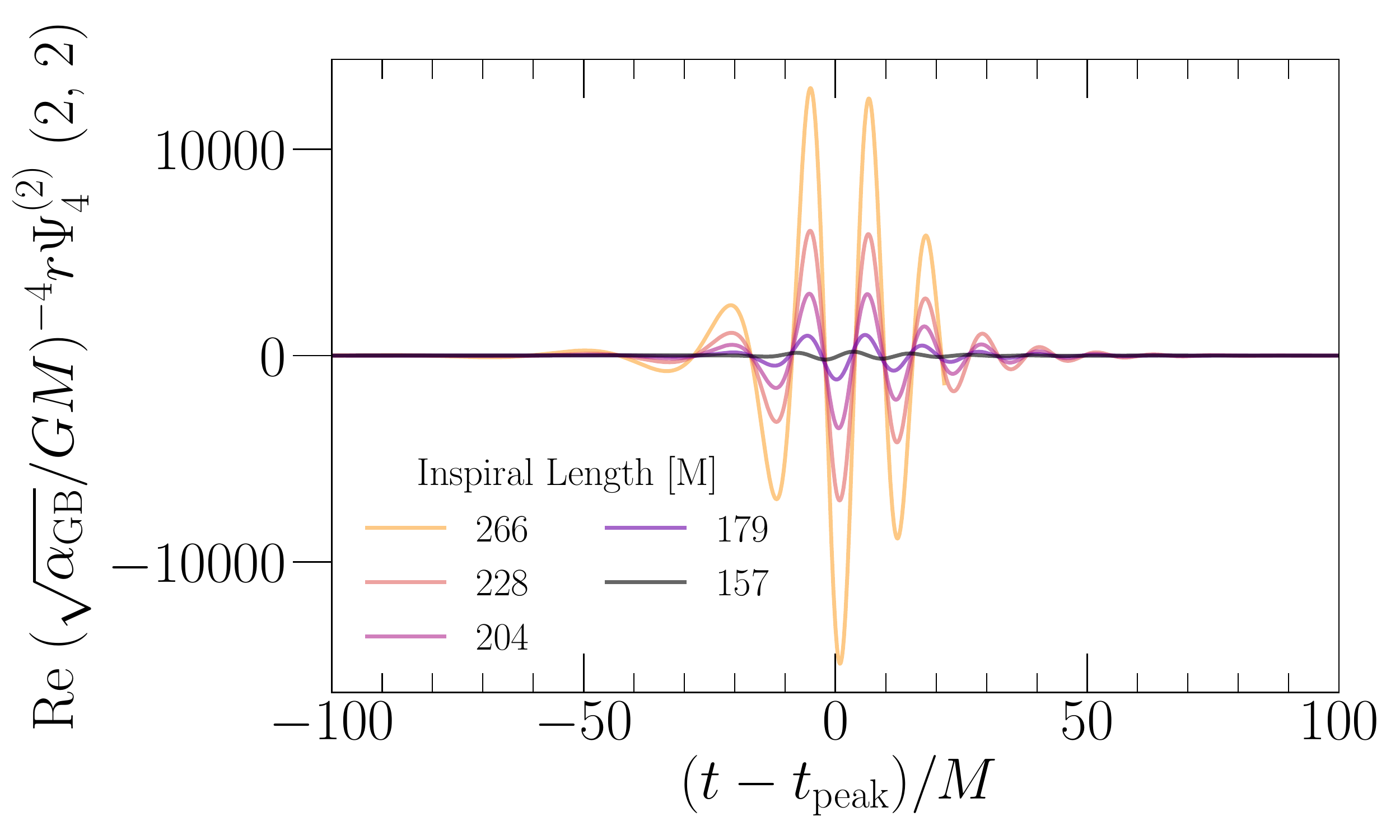}
  \caption{Secular growth in leading-order EDGB gravitational waveforms as function of inspiral length of the waveform. Each colored curve corresponds to a simulation with a different start time for the EDGB fields (as discussed in Sec.~\ref{sec:secular_growth}), with the same GR background simulation for each. We label each curve by the time difference between the peak of the waveform and the start time of ramping on the EDGB field (minus the ramp time). We see that simulations with earlier EDGB start times have higher amplitudes at merger, having had more time to accumulate secular growth.}
  \label{fig:SecularhPsi4}
\end{figure}

\begin{figure}
  \includegraphics[width=\columnwidth]{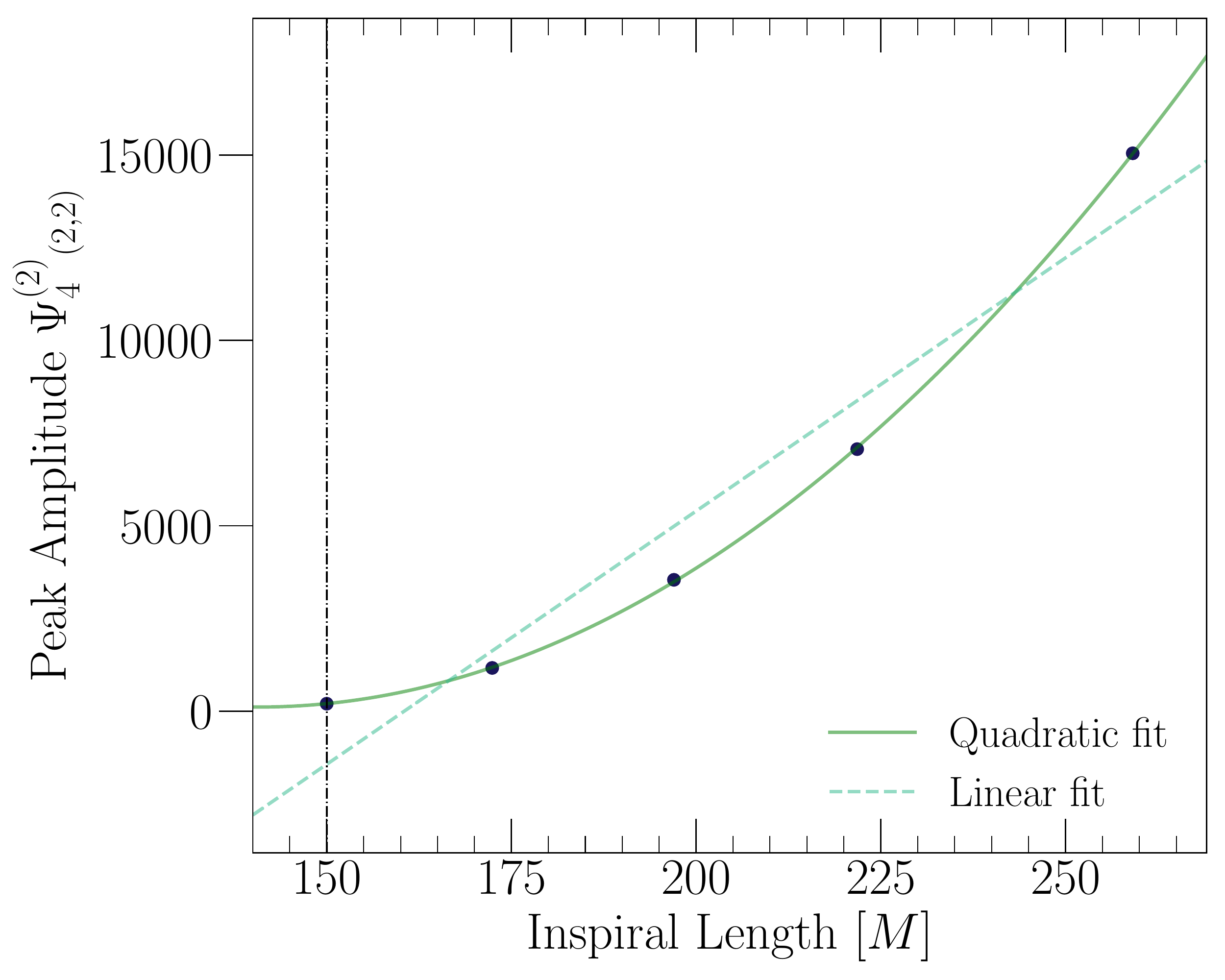}
  \caption{Peak amplitude of the EDGB correction to the gravitational waveform as a function of inspiral length. We show the length relative to the peak of the waveform (as in Fig.~\ref{fig:SecularhPsi4}). The dashed black vertical line corresponds to the length of the EDGB merger simulation we present in this paper. The peak amplitude serves as a measure of the amount of secular growth in the waveform (cf. Fig.~\ref{fig:SecularhPsi4}). We see that the secular growth attains a quadratic minimum, and thus for a short enough inspiral length, we can obtain an EDGB gravitational waveform with minimal secular contamination.
  }
  \label{fig:PeakAmplitude}
\end{figure}

As discussed in Sec.~\ref{sec:secular_growth}, the perturbative scheme leads to secular growth in the inspiral waveform. In Fig.~\ref{fig:SecularhPsi4}, we show the leading-order EDGB correction to the gravitational waveform for a variety of simulation lengths (with the same background GR simulation). We ramp on the EDGB source terms at different start times in order to produce different inspiral lengths, as discussed in Sec.~\ref{sec:secular_growth}. We see that the longest simulations have the largest amplitude at merger, consistent with secular growth. In Fig.~\ref{fig:PeakAmplitude}, we take a more quantitative look, plotting the peak amplitude of the waveform as a function of inspiral length. In the dCS case (cf. Fig. 7 of~\cite{MashadCSGWPaper}), we saw that for the closest start time to merger, the secular growth attained a quadratic minimum. In other words, the merger waveform we presented was not contaminated by secular effects. 

In Fig.~\ref{fig:PeakAmplitude}, we see a similar quadratic minimum for the EDGB correction to the waveform, although this occurs at a shorter inspiral length (later start time) than in dCS. This higher level of secular growth in EDGB than in dCS is consistent with the theoretical predictions of Sec.~\ref{sec:secular_growth}, as the EDGB inspiral is more heavily modified than in dCS due to the presence of dipolar radiation~\cite{Yagi:2011xp}. 

\section{Constraints on $\sagb$ from EdGB merger waveforms}
\label{sec:analysis} 

As shown in Sec.~\ref{sec:merger_waveform}, we have access to the leading-order EDGB merger waveform for a GW150914-like system. What sort of physical constraints on EDGB can we extract from the merger phase? 

\subsection{Merger mismatches}

The first step that we can take is to perform a merger-only analysis by computing mismatches between the GR waveform and the EDGB waveform using the formulae in Sec.~\ref{sec:mismatch_appendix}. This involves restricting to a given time (or frequency) range over which to compute the mismatch. When performing tests of general relativity, LIGO performs such merger-only calculations. In~\cite{TheLIGOScientific:2016src}, the authors performed an inspiral-merger-ringdown consistency test for GW150914 by inferring final mass and spin parameters using GR waveforms from the post-inspiral portion of the waveform only, from the inspiral portion of the waveform only, and comparing the resulting posterior distribution to that from the full waveform analysis. For GW150914, the merger-ringdown region was chosen to be $[132, 1024]$ Hz. In this region, the signal had a signal to noise ratio (SNR) of 16, which is larger than the full-waveform SNR of the other nine BBH detections in GWTC-1~\cite{LIGOScientific:2018mvr}. 

\begin{figure}
  \includegraphics[width=\columnwidth]{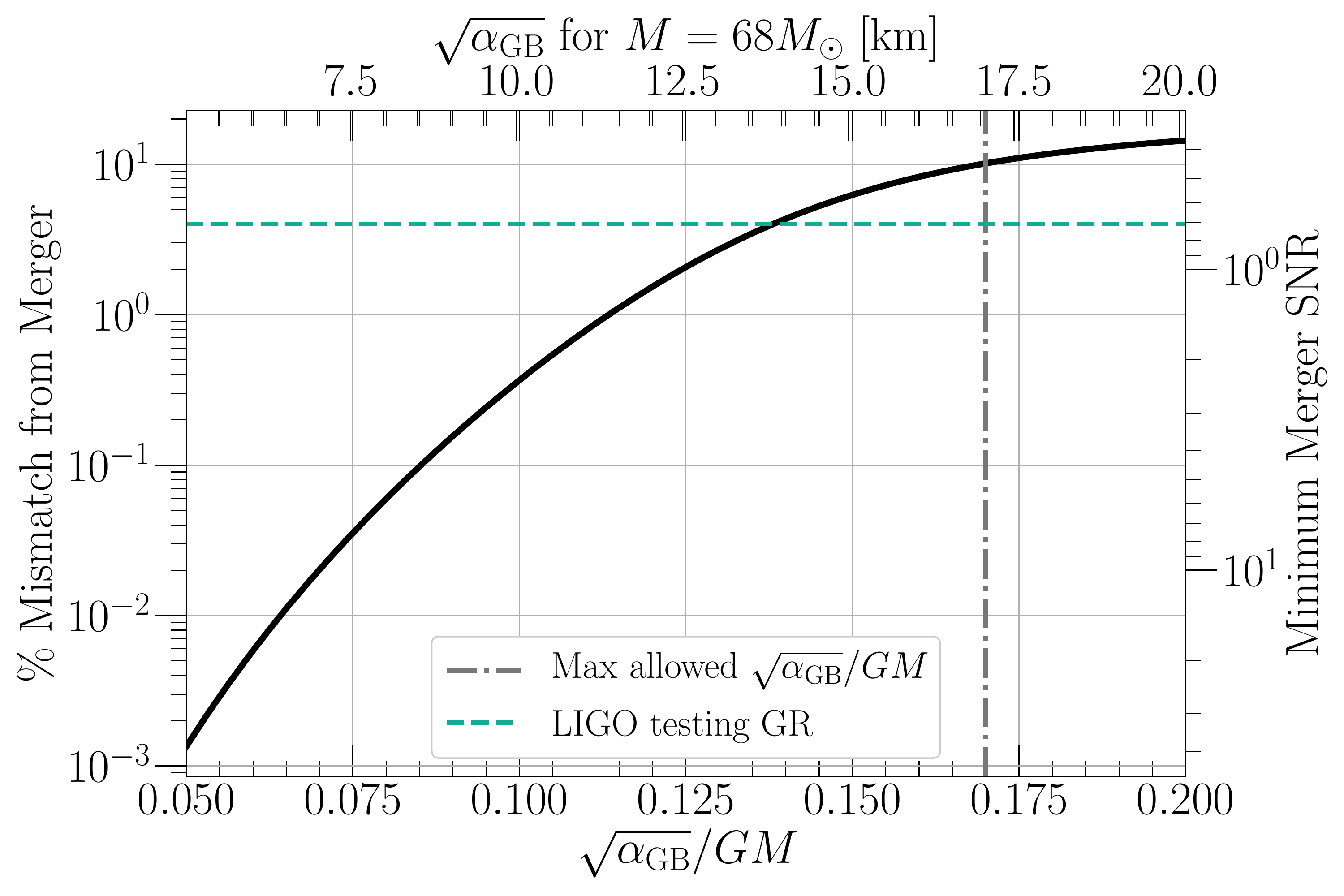}
  \caption{Mismatch between general relativity GW150914 waveform (cf. Sec.~\ref{sec:simulation}) and the corresponding EDGB-corrected gravitational waveform, as defined in Eq.~\eqref{eq:mismatch}. We show the mismatch for our merger waveform as a function of the EDGB coupling parameter, $\sgmagb$. We show the maximum allowed value of $\sgmagb$ from the regime of validity (cf. Sec.~\ref{sec:validity}) in dot-dashed gray. The dashed horizontal line corresponds to the LIGO mismatch of $4\%$ from testing GR with GW150914~\cite{TheLIGOScientific:2016src}. The top vertical axis corresponds to $\sagb$ computed from $\sgmagb$ on the bottom axis assuming that $M = 68 M_\odot$ for GW150914. 
  }
  \label{fig:Mismatch}
\end{figure}

We thus compute mismatches between the GR and EDGB merger waveforms, shown in Fig.~\ref{fig:Mismatch}. We show the mismatch (cf. Sec.~\ref{sec:mismatch_appendix}) for various values of $\sgmagb$ (cf. Fig.~\ref{fig:Together}). In particular, for a $1\%$ mismatch, we find $\sgmagb \lesssim 0.11$. For GW150914, we choose $M \sim 68\, M_\odot$~\cite{TheLIGOScientific:2016wfe}, and thus compute $\sagb \lesssim 11$ km.
Note that though we shift the waveforms in time and phase to compute a minimum mismatch, we do not vary the GR waveform parameters (mass and spin). Thus our mismatch estimate is optimistic, and performing a full parameter-estimation analysis on our EDGB waveform is the subject of future research. 

For heavier BBH systems, such as GW170729 with $M = 84\,M_\odot$, which had 3 cycles in the LIGO band~\cite{LIGOScientific:2018mvr}, we can in theory use only the merger-ringdown EDGB waveforms from numerical relativity simulations for data analysis, without requiring EDGB inspiral waveforms. Note, however, that with all other parameters held equal, this lead to a lower constraint on $\sagb$ from the larger total mass. Moreover, GW170729 has an SNR of $\sim 10$, which is less than the merger SNR of 16 for GW150914. Note that the LIGO intermediate mass black hole search~\cite{Salemi:2019ovz} which looked for BBHs with $M \in [120, 800]\,M_\odot$ did not detect any signals. 

\subsection{Including inspiral}

How much more could we gain if we additionally included the inspiral phase? Gaining access to the inspiral phase for EDGB waveforms is ongoing work, through either implementing a renormalization scheme to remove secular effects as outlined in~\cite{MashadCSGWPaper}, or by stitching on post-Newtonian or parametrized post-Einstenian (ppE) EDGB waveforms for the inspiral~\cite{Yagi:2011xp, Tahura:2018zuq}, to obtain a full waveform. 

In~\cite{Yunes:2016jcc}, the authors use the ppE formalism to bound $\sagb$ with GW150914. Fig. 15 of~\cite{Yunes:2016jcc} shows the upper bounds on $\sagb$, including values of $\mathcal{O}(20, 40)$ km, but this is very sensitive to the dimensionless spins of the black holes, which are poorly constrained (cf.~\cite{TheLIGOScientific:2016wfe, LIGOScientific:2018mvr}). Thus, the authors do not place an upper bound on $\sagb$. In~\cite{Tahura:2019dgr}, the authors place an upper bound of $\sagb \lesssim 51.5$ km for GW150914 using a ppE analysis, which is higher than our merger-only analysis bound. Including a merger phase to these inspiral-only analyses can thus improve their bounds on $\sagb$.

\subsection{Comparison to observational and projected constraints}

Let us now compare the merger-analysis result of $\sagb \lesssim 11$ km with \textit{observational} and predicted observational constraints in the literature. We summarize these present constraints in Table~\ref{tab:bounds}. Most notably, Witek et al.~\cite{Witek:2018dmd} estimate from their scalar field calculations that for a GW151226-like system~\cite{Abbott:2016nmj}, the constraint would be $\sagb \lesssim 2.7$ km. Note that this signal has $\sim 15$ cycles in the LIGO band (compared to $\sim 5$ in the LIGO band for GW150914)~\cite{LIGOScientific:2018mvr}, and thus the inspiral phase, which is not included in our estimate, plays a greater role for this system. Moreover, this esimate was performed with a mass ratio of $q \sim 2$ and total mass $\sim 20 M_\odot$, which leads to stronger beyond-GR effects due to the higher curvature of the smaller object. 

\begin{table}[htb!]
\begin{center}
  \begin{tabular}{ l | l }
    Reference & $\sagb$ bound \\ \hline \hline
    \textbf{Cassini Shapiro time delay}\cite{2003Natur.425..374B} & $\lesssim \mathcal{O}(10^7)$ km \\
    \textbf{X-ray binary orbital decay}~\cite{Yagi:2012gp} & $\lesssim 10$ km \\
    Compact star stability~\cite{Pani:2011xm} & $ \lesssim 5.4$ km \\
    LIGO SNR 30 detections~\cite{Stein:2013wza} & $\lesssim \mathcal{O}(1-10)$ km \\
    EDGB scalar simulations for GW151226~\cite{Witek:2018dmd} & $\lesssim 2.7$ km \\
    GW150914 ppE~\cite{Tahura:2019dgr} & $\lesssim 51.5$ km 
  \end{tabular}
\caption{Observed and projected bounds on the EDGB lengthscale from various studies. The first to rows (in bold), correspond to observed bounds, from the Cassini probe constraints on Shapiro time delay and observations of X-ray binaries. Note that all bounds are given in terms of the conventions in our action (cf. Eq.~\eqref{eq:action}), chosen to be consistent with~\cite{Witek:2018dmd}.
}
\label{tab:bounds}
\end{center}
\end{table}

\section{Conclusion}
\label{sec:conclusion}

We have produced the first astrophysically-relevant numerical relativity binary black hole gravitational waveform in Einstein dilaton Gauss-Bonnet gravity, a beyond-GR theory of gravity. We have focused on a system with parameters consistent with GW150914, the loudest LIGO detection thus far. This extends our previous work for producing such a waveform for GW150914 in dynamical Chern-Simons gravity~\cite{MashadCSGWPaper}. 

In Sec.~\ref{sec:setup}, we laid out our order-reduction scheme, which we use to obtain a well-posed initial value formulation and produce the leading-order EDGB correction to the gravitational waveform. In Sec.~\ref{sec:merger_waveform}, we showed the EDGB-corrected waveforms for a system consistent with GW150914 (cf. Sec.~\ref{sec:simulation}). We find that there is secular growth in the inspiral phase (Sec.~\ref{sec:observed_secular}), and thus present a merger-ringdown waveform that is free of secular growth. 

We thus focus on a post-inspiral-only analysis, and compute the mismatch between the (background) GR waveform and the EDGB-corrected waveforms, finding a bound on the EDGB coupling parameter of $\sgmagb \lesssim 11$ km. This is a stronger result than inspiral-only analyses for GW150914, which bound $\sagb \lesssim 51.5$ km. Note that GW150914 has an SNR of 16 in the post-inspiral phase (cf.~\cite{TheLIGOScientific:2016src}), which is larger than the total SNR of each other event in GWTC-1~\cite{LIGOScientific:2018mvr}. Stitching on a parametrized post-Einstenian EDGB inspiral or removing the inspiral secular growth from our simulations (cf.~\cite{MashadCSGWPaper}) to take full advantage of an inspiral-merger-ringdown analysis is the subject of future work.

Our ultimate goal is to make these beyond-GR waveforms useful for LIGO
and Virgo tests of general relativity~\cite{TheLIGOScientific:2016src,
  LIGOScientific:2019fpa}. We can improve the mismatch analysis  by allowing the GR waveform parameters to vary, thus checking for degeneracies in the GR-EDGB parameter space. Moreover, we can perform a more quantitative analysis by injecting our beyond-GR waveforms
into LIGO noise and computing posteriors recovered using present LIGO parameter estimation and testing-GR methods~\cite{Aasi:2013jjl, Cornish:2014kda,
  TheLIGOScientific:2016wfe, LIGOScientific:2019fpa}. Ultimately, we would like to generate enough beyond-GR EDGB waveforms to fill the BBH parameter space. We can then produce a beyond-GR surrogate model~\cite{Varma:2019csw} and perform model-dependent tests of GR.

\section*{Acknowledgements}

We thank Leo Stein, Helvi Witek, and Paolo Pani for useful discussions. The Flatiron Institute is supported by the Simons Foundation. Computations were performed using the Spectral Einstein Code~\cite{SpECwebsite}. All computations were performed on the Wheeler cluster at Caltech, which is supported by the Sherman Fairchild Foundation and by Caltech.

\appendix

\section{Mismatches}
\label{sec:mismatch_appendix}

Given the GR and EDGB-corrected waveforms (as shown in Fig.~\ref{fig:Together}), let us consider the mismatch between these waveforms. A more involved calculation would involve computing a mismatch in the presence of gravitational wave detector noise and considering a range of parameters for the GR waveform to test for degeneracies~\cite{Chatziioannou:2017tdw}. Here, we perform a simpler mismatch calculation between the background GR waveform $\Psi_4^{(0)}$ and the corresponding EDGB-modified waveform considered in this study (cf. Sec.~\ref{sec:simulation}). Once we have the EDGB correction $\Psi_4^{(2)}$ from the numerical relativity simulation, we introduce a coupling parameter $\sgmagb$ before adding it to the GR waveform using Eq.~\eqref{eq:total_psi4} to obtain $\Psi_4(\sagb)$. 

We then compute the mismatch as (cf.~\cite{Boyle:2019kee})
\begin{align}
\label{eq:mismatch}
    &\mathrm{Mismatch}(\sagb) \equiv \\
    &\nn \quad 1 - \mathrm{Re} \left(\frac{\langle \Psi_4^{(0)}, \Psi_4 (\sagb)\rangle }{\sqrt{\langle \Psi_4^{(0)}, \Psi_4^{(0)} \rangle \times \langle \Psi_4 (\sagb) ,\Psi_4 (\sagb)\rangle }} \right)\,,
\end{align}
where we have explicitly shown the dependence on $\sagb$. We define the inner product $\langle\,,\,\rangle$ between two waveforms as
\begin{align}
\label{eq:inner_product}
    \langle \Psi_4 {}^{[1]}, \Psi_4 {}^{[2]} \rangle \equiv \int_{t_\mathrm{start}}^{t_\mathrm{end}} \Psi_4 (t) {}^{[2]} \tilde \Psi_4^* (t) {}^{[1]} dt\,,
\end{align}
where ${}^*$ denotes complex conjugation. This is precisely the inner product used in~\cite{Boyle:2019kee}. We choose $t_\mathrm{start}$ to be the section of the waveform where EDGB effects are fully ramped-on, and choose $t_\mathrm{end}$ to be the end of the numerical waveform. This is equivalent, by Parseval's theorem, to a noise-weighted inner product in the frequency domain with noise power spectral density $S_n(|f|) = 1$. We shift the waveforms in time and phase when computing this overlap.

%%%%%%%%%%%%%%%%%%%%%%%%%%%%%%%%%%%%%%%%%%%%%%%%%%%%%%%%%%%%%%%%%%%%%%%%%%%%%%%
%% BIBLIOGRAPHY
%%%%%%%%%%%%%%%%%%%%%%%%%%%%%%%%%%%%%%%%%%%%%%%%%%%%%%%%%%%%%%%%%%%%%%%%%%%%%%%
\bibliography{biblio}
\end{document}